# Clarifying Selection Bias in Cluster Randomized Trials

**Authors:** Fan Li[1] (PhD), Zizhong Tian (MS), Jennifer Bobb (PhD), Georgia Papadogeorgou (PhD), Fan Li[2] (PhD)

**Correspondence to** Fan Li[1], Department of Statistical Science, Duke University, Durham, North Carolina (email: fl35@duke.edu)

**Author affiliations:**

Department of Statistical Science, Duke University, Durham, North Carolina (Fan Li[1], PhD); Department of Public Health Sciences, Pennsylvania State University, Hershey, Pennsylvania (Zizhong Tian, MS); Kaiser Permanente Washington Health Research Institute, and Department of Biostatistics, University of Washington, Seattle, Washington (Jennifer Bobb, PhD); Department of Statistics, University of Florida, Gainesville, Florida (Georgia Papadogeorgou, PhD); Department of Biostatistics, Yale University School of Public Health, New Haven, Connecticut (Fan Li[2], PhD)

**Conflict of interest:** none declared.

**Funding:** This research is supported in part by the Patient-Centered Outcomes Research Institute (PCORI) contract ME-2019C1-16146. The contents of this article are solely the responsibility of the authors and do not necessarily represent the view of PCORI.

**Data availability statement:** This paper does not use real data.

**Running head:** Selection bias in cluster randomized trials

**Word Count:** (Abstract 370); (Main text 3998)

**Abbreviations:**

ARTEMIS: Affordability and Real-World Antiplatelet Treatment Effectiveness After Myocardial Infarction Study

ATE: average treatment effect

ITT: intention-to-treat

**Abstract**

**Background and objective**: In cluster randomized trials, patients are typically recruited after clusters are randomized, and the recruiters and patients may not be blinded to the assignment. This often leads to differential recruitment and consequently systematic differences in baseline characteristics of the recruited patients between intervention and control arms, inducing post-randomization selection bias. We aim to rigorously define causal estimands in the presence of selection bias. We elucidate the conditions under which standard covariate adjustment methods can validly estimate these estimands. We further discuss the additional data and assumptions necessary for estimating causal effects when such conditions are not met.

**Methods**: Adopting the principal stratification framework in causal inference, we clarify there are two average treatment effect (ATE) estimands in cluster randomized trials: one for the overall population and one for the recruited population. We derive the analytical formula of the two estimands in terms of principal-stratum-specific causal effects. Further, using simulation studies, we assess the empirical performance of the multivariable regression adjustment method under different data generating processes leading to selection bias.

**Results**: When treatment effects are heterogeneous across principal strata, the ATE on the overall population generally differs from the ATE on the recruited population. A naïve intention-to-treat analysis of the recruited sample leads to biased estimates of both ATEs. In the presence of post-randomization selection and without additional data on the non-recruited subjects, the ATE on the recruited population is estimable only when the treatment effects are homogenous between principal strata, and the ATE on the overall population is generally not estimable. The extent to which covariate adjustment can remove selection bias depends on the degree of effect heterogeneity across principal strata.

**Conclusion**: There is a need and opportunity to improve the analysis of cluster randomized trials that are subject to post-randomization selection bias. For studies prone to selection bias, it is important to explicitly specify the target population that the causal estimands are defined on and adopt design and estimation strategies accordingly. To draw valid inferences about treatment effects, investigators should (i) assess the possibility of heterogeneous treatment effects, and (ii) consider collecting data on covariates that are predictive of the recruitment process, and on the non-recruited population from external sources such as electronic health records.

**Keywords**: average treatment effect, causal inference, heterogeneous treatment effect, intention-to-treat, recruitment bias, identification bias, principal stratification

**Background**

In cluster randomized trials, treatment is randomly assigned at cluster level, all individuals in a cluster receive the same treatment, and outcomes are typically measured at the individual level. Cluster randomized trials are often used to study interventions that are impractical to be assigned to individuals; they are also advocated to minimize treatment "contamination" between intervention and control participants.[1–3] This design has been increasingly popular in pragmatic trials for comparative effectiveness research. Compared to traditional individually randomized trials, a main challenge in cluster randomized trial is the potential for post-randomization selection bias.[4] Specifically, subjects are usually recruited after clusters are randomized, but both the recruiters and subjects are not blinded to the randomized treatment assignment.[5,6] The assignment can therefore affect the recruitment process, leading to differential recruitment in intervention and control clusters and consequently systematic differences between the subjects in the two arms.[7,8] This problem is particularly common when the invention is perceived to be beneficial or disadvantageous by patients. One example is the ARTEMIS (Affordability and Real-World Antiplatelet Treatment Effectiveness After Myocardial Infarction Study), a pragmatic trial designed to determine whether removing co-payment barriers increases P2Y12 inhibitor persistence and lowers risk of major adverse cardiovascular events among patients who had acute myocardial infarction.[9] In ARTEMIS, hospitals randomized to the intervention arm provided recruited patients with co-payment vouchers for clopidogrel or ticagrelor for 1 year, and hospitals randomized to the control arm did not provide vouchers. The financial incentive renders a much higher recruitment rate in the intervention arm than in the control arm, and significant imbalances in many individual-level baseline covariates among the recruited patients.

Selection bias can also arise even when there is no participant recruitment, because participants often need to be prospectively identified after cluster randomization. Here, we use the term, "recruitment", to generically refer to the inclusion into a study. Regardless of the context of "recruitment", the common nature of the aforementioned selection bias is that the identification of trial participants (either through formal recruitment or using existing data sources) occurs after randomization and is partially driven by the cluster assignment. Such post-randomization selection breaks the initial randomization. As Hernan and Robins (2020, p 103)[10] put: "randomization protects against confounding, but not against selection bias when the selection occurs after the randomization." This type of selection bias has been known as recruitment bias or identification bias in the literature.[11,12] The best way to avoid selection bias is through careful design.[5] But despite such efforts in the design stage, selection bias can still persist in cluster randomized trials. In these situations, a common practice is to combine intention-to-treat (ITT) analysis with covariate adjustment, via regression adjustment or propensity scores,[13,14] in the analysis stage. However, theoretical justification for these methods in this setting is not clear because they are designed to correct for chance imbalance rather than post-randomization selection bias. More generally, there is a lack of rigorous discussion of causal estimands, as well as design and analysis strategies to address selection bias in cluster randomized trials. In this paper, we investigate this problem using principal stratification,[15] which is a general framework for addressing post-treatment confounding in causal inference. We clarify different target populations, define corresponding causal estimands, and illustrate the implications of post-randomization confounding. Using analytical derivations and simulations, we demonstrate that when heterogeneous treatment effects are present: *(i)* the average treatment effect (ATE) on the overall population is different from the ATE on the recruited population, *(ii)* a naive ITT analysis on the recruited sample can be biased for both ATE estimands, and *(iii)* standard covariate adjustment methods alone are often not adequate to correct for post-randomization selection bias. Furthermore, we discuss the additional data and assumptions that are necessary for unbiased estimation of the causal estimands in such situations. Note that post-randomization selection bias is not specific to cluster randomized trials and our following discussion is also applicable to individual-level randomized trials. However, because cluster trials are particularly prone to such selection bias, we will focus on this context.

**Methods**

This section introduces the study setup, notations, and estimands. Assume we have $I$ clusters, $m$ of which are randomized to the intervention arm, denoted by $Z_i = 1$, and the remaining clusters to the control arm, denoted by $Z_i = 0$ ($i = 1, 2, \ldots, I$). We assume each subject $j$ in cluster $i$ has a pair of potential outcomes corresponding to intervention and control, $\{Y_{ij}(1),\ Y_{ij}(0)\}$, of which only the one under the assigned arm is observed, denoted by $Y_{ij} = Y_{ij}(Z_i)$. In a cluster randomized trial, subjects are often recruited after the cluster treatment assignment, and therefore not all subjects present in a cluster are recruited into the study. For each subject $j$ in cluster $i$, we define a recruitment status $R_{ij}$ such that $R_{ij} = 1$ if the subject is recruited into the trial and $R_{ij} = 0$ if not. Denote the number of recruited subjects in cluster $i$ as $N_i$, and the total sample size of the trial as $N = \sum_{i=1}^{I} N_i$. Each subject also has a set of baseline covariates $X_{ij}$. Usually, we observe the outcome and covariates only for the recruited subjects, and this is the scenario we consider here.

As a result of recruitment, there are two different causal estimands. The first is the average treatment effect (ATE) on the overall population, corresponding to all subjects in the study clusters, recruited or not:

$$\tau^O = \mathbb{E}[Y_{ij}(1) - Y_{ij}(0)].$$

A second estimand is the ATE on the recruited population, defined only on the recruited subjects:

$$\tau^R = \mathbb{E}[Y_{ij}(1) - Y_{ij}(0)|R_{ij} = 1].$$

The estimand $\tau^R$ is also commonly known as the ITT effect in clinical trials. The expectation here is over the super population of clusters and units. For more technical discussion of the estimands, see Su and Ding.[16] Usually the intended target population is either the entire or a pre-specified subset (e.g. patients identified with a certain medical condition) of the overall population, and thus $\tau^O$ is the intended target estimand. The two estimands $\tau^O$ and $\tau^R$ are identical if the recruited population is a simple random sample of the overall population and/or the treatment effect is homogenous across subjects. However, neither condition is generally true. In fact, because $\tau^R$ is defined conditional on a *post*-randomization variable $R_{ij}$ that can be affected by the treatment assignment, it is usually different from $\tau^O$ when there is treatment effect heterogeneity. Randomization ensures the intervention and control arms are comparable in the overall population, eliminating all the selection bias that could occur *before* the assignment. But it does not give the same guarantee if the post-randomization recruitment process depends on covariates, which renders (i) the recruited sample not representative of the overall population, and (ii) the recruited treated and control groups imbalanced in their characteristics. Below we adopt the principal stratification framework[15]—a generalization of the instrumental variable approach to noncompliance in randomized experiments—to illustrate post-randomization selection in the recruitment process.

Because recruitment occurred after randomization, we can consider that each subject also has two potential recruitment statuses corresponding to intervention and control, $\{R_{ij}(1), R_{ij}(0)\}$. This allows us to cross-classify subjects in the overall population into different principal strata, that is, the joint potential recruitment values under intervention and control, $S_{ij} = (R_{ij}(1), R_{ij}(0))$. Specifically, there are four principal strata: always-recruited, $S_{ij} = (1,1) = a$, subjects who would be recruited irrespective of their cluster's treatment assignment; never-recruited, $S_{ij} = (0,0) = n$, subjects who would not be recruited irrespective of their assignment; compliant-recruited, $S_{ij} = (1,0) = c$, subjects who would be recruited if assigned to intervention arm, but would not if to control; defiant-recruited, $S_{ij} = (0,1) = d$, subjects who would be recruited if assigned to control arm, but would not if to intervention. The above nomenclature originates from the instrumental variable literature[17] to noncompliance where the randomized assignment is viewed as an instrument.

The central property of principal strata is that, by definition, each subject's stratum membership is not affected by the assignment, and thus is a pre-randomization variable. Then, the principal causal effects are defined as the causal effects within each principal stratum:

$$\tau_s = \mathbb{E}[Y_{ij}(1) - Y_{ij}(0) | S_{ij} = s], \text{ for all } s \in \mathbb{S} = \{a, n, c, d\}.$$

It is easy to show that the overall ATE is a weighted average of the principal causal effects:

$$\tau^O = \sum_{s \in \mathbb{S}} \tau_s \, p_s, \tag{1}$$

where $p_s = P(S_{ij} = s)$ is the proportion of stratum $s$ in the overall population. When the recruited population is a representative sample of the overall population, $\tau^R = \tau^O$. Also, when the treatment effect is homogenous across all subjects, all the estimands are equal: $\tau^R = \tau^O = \tau_s$. However, this equality does not hold generally. In the next section, we will use analytical derivation and numerical simulations to demonstrate that heterogeneity across principal strata leads to systematic differences between $\tau^R$ and $\tau^O$ in the presence of post-randomization selection, and standard covariate adjustment methods cannot eliminate such selection bias

**Results**

*Analytical Derivation*

This subsection analytically illustrates the implications of post-randomization selection in a simplified scenario without baseline covariates. Recall that principal strata are defined as the joint potential recruitment status under both treatment values, only one of which is observed. Therefore, the individual stratum membership is not directly observed, and we need to make some additional assumptions to estimate the principal causal effects.

We maintain two assumptions. The first assumption is *cluster randomization*, with the randomization probability $r = P(Z_i = 1)$ between 0 and 1. This assumption ensures that the assignment of clusters to intervention or control does not depend on any covariates or outcomes. The second assumption is *monotonicity*, which states that $R_{ij}(1) \geq R_{ij}(0)$ for each subject $j$ in cluster $i$. Monotonicity requires that a patient who would be recruited in the control arm would also be recruited in the intervention arm, but not vice versa; this assumption rules out defiant-recruited patients. Monotonicity is standard in the literature and is plausible in many real applications. For example, in ARTEMIS, because the intervention reduced copayment of patients, it was obviously more attractive to the patients than the control condition. Hence, it is reasonable to assume that the patients who would be recruited under the control would also be recruited under the intervention, but not vice versa.

By definition, the recruited population ($R_{ij} = 1$) does not consist of any never-recruited subjects. So, combined with monotonicity, the recruited population only consists of always-recruited and compliant-recruited subjects. Furthermore, under monotonicity, the recruited subjects in the intervention arm (i.e., $Z_i = 1$, $R_{ij}(1) = 1$) can be either always-recruited or compliant-recruited, whereas the recruited patients in the control arm (i.e., $Z_i = 0$, $R_{ij}(0) = 1$) can only be always-recruited. If the average treatment effects among the always-recruited and compliant-recruited patients are heterogeneous, which is the case in most real-world situations, then randomization no longer holds among the recruited population. This has several important implications for treatment effect estimation.

First, we can analytically express $\tau^R$ in terms of the principal causal effects, as follows.

*Result 1. Assuming random assignment and monotonicity, the ATE on the recruited population is*

$$\tau^R = \frac{rp_c}{rp_c + p_a} \tau_c + \left(1 - \frac{rp_c}{rp_c + p_a}\right) \tau_a. \qquad (2)$$

The proof is given in Appendix A. Result 1 show that the estimand $\tau^R$ is a weighted average of the treatment effect in the always-recruited and the compliant-recruited strata, and it depends on the ratio between the proportions of the two strata. Comparing formula (1) and (2), we can show that $\tau^R \neq \tau^O$ unless the treatment effect is homogenous across principal strata, i.e., $\tau_c = \tau_a = \tau_n$. Result 1 also shows that $\tau^R$ depends on the randomization probability of a trial, and thus it can vary for the same overall population depending on the cluster allocation proportion. This suggests that while $\tau^R$ is a valid causal estimand, its interpretation is specific to each trial with a certain randomization probability.

Second, the different composition of strata between the arms in the recruited population implies that these two arms are no longer comparable. For example, in ARTEMIS, patients who would be recruited regardless of the random assignment may be more aware of health information and thus more supportive of clinical research than patients who would be recruited only when assigned to the intervention, or differ in other ways, both measured and unmeasured. Consequently, recruited patients in the intervention arm (which includes always-recruited and compliant-recruited patients) differ systematically from recruited patients in the control arm (which includes always-recruited patients only). Analytically, this implies that a standard *ITT analysis of the recruited population*, i.e., the difference of the averaged outcomes between the arms,

$$\hat{\tau}^{ITT} = \frac{\sum_{Z_i=1} R_{ij} Y_{ij}}{\sum_{Z_i=1} R_{ij}} - \frac{\sum_{Z_i=0} R_{ij} Y_{ij}}{\sum_{Z_i=0} R_{ij}},$$

generally leads to a biased estimate of both $\tau^O$ and $\tau^R$.

To illustrate this point numerically, we provide a simple hypothetical example without covariates in Figure 1. In this example, we assume (i) monotonicity, (ii) equal proportion of each stratum in the overall population (i.e. $p_c = p_a = p_n = 1/3$), and (iii) treatment effects that are heterogeneous across strata, with the true effects given as $\tau^O = 15$, $\tau_a = 20$, $\tau_c = 15$, $\tau_n = 10$. Heterogeneous treatment effects are highly plausible given expected differences in always-recruited and compliant-recruited patients. An ITT analysis of the recruited sample gives an estimate of $\hat{\tau}^{ITT} = 17.5$, which is biased for the true $\tau^O$. Moreover, according to formula (2), $\hat{\tau}^{ITT}$ is also biased for $\tau^R$ for any $0 < r < 1$. Such bias arises because post-randomization selection breaks the initial randomization, and the recruited treated and control patients are composed of different principal strata. Because an individual's principal stratum is not directly observed, this induces a type of post-randomization unmeasured confounding.

[Figure 1 about here]

Third, besides heterogeneity in outcomes and treatment effects, subjects usually also differ in baseline covariates across principal strata. Therefore, systematic imbalance in observed covariates is expected between the recruited patients in the two arms when post-randomization selection occurs. The underlying mechanism causing such imbalance is distinct from that causing chance imbalance;[18,19] this has important practical implications for analysis. Specifically, in the randomized trial literature, covariate adjustment methods such as regression adjustment or propensity scores are often adopted to improve precision of effect estimation by accounting for chance imbalances in baseline covariates.[19] However, these methods are not designed to correct for post-randomization selection bias, and thus applying them to the recruited sample in a cluster randomized trial would generally lead to biased causal estimates except for certain specific situations. Additional data and assumptions are necessary for valid causal inference. We will further demonstrate this point in the next section.

*Simulation Studies*

This subsection carries out simulations to illustrate that covariate adjustment may not be adequate to address post-randomization selection bias in cluster randomized trials. Based solely on the observed (recruited) sample, we can at most estimate the ATE on the recruited population $\tau^R$, and thus below we will focus on $\tau^R$ and leave the discussion of $\tau^O$ to the concluding section.

Assume there are $I = 20$ clusters participating in a cluster randomized trial, half of which are randomized to the intervention arm. We assume that each cluster consists of 500 subjects, and the total overall population size is 10,000. We simulate one continuous and one binary covariate for each member of the total population: $X_{1ij} \sim N(\mu_i, 1)$ where the cluster-specific mean $\mu_i \sim N(0,1)$ and $X_{2ij} \sim Bernoulli(0.4)$; we denote $X_{ij} = (X_{1ij}, X_{2ij})'$. The latent principal stratum membership for each member in the overall population is generated from a multinomial logistic model with probabilities:

$$P(S_{ij} = s|X_{ij}) = \frac{\mathbb{I}(s = a) \exp(\beta_{a0} + \beta'_{a1}X_{ij}) + \mathbb{I}(s = c) \exp(\beta_{c0} + \beta'_{c1}X_{ij}) + \mathbb{I}(s = n)}{\exp(\beta_{a0} + \beta'_{a1}X_{ij}) + \exp(\beta_{c0} + \beta'_{c1}X_{ij}) + 1}$$

for $s = a, c, n$. We set the parameter $(\beta_{a0}, \beta'_{a1}) = (0.3, 0.2, 0.1)$ and $(\beta_{c0}, \beta'_{c1}) = (0.1, 0.2, -0.1)$ such that the marginal population stratum proportions are $(p_n, p_a, p_c) \approx (0.3, 0.4, 0.3)$.

As discussed earlier, typically only a subset of the subjects in each cluster are recruited in a study. We mimic this realistic setting in the simulations: we assume that each cluster aims to recruit 50 patients (out of 500 patients) so that the total trial sample size is 1000. However, due to post-randomization selection, each intervention cluster will only recruit from the always-recruited $(S_{ij} = a)$ or the compliant-recruited $(S_{ij} = c)$ in that cluster, whereas each control cluster will only recruit from the always-recruited $(S_{ij} = a)$. For the recruited subjects, we simulate the potential outcomes from a linear mixed model,

$$Y_{ij}(z) = I(S_{ij} = a)\left(\mu_a + \tau_a z + \lambda'_a X_{ij}\right) + I(S_{ij} = c)\left(\mu_c + \tau_c z + \lambda'_c X_{ij}\right) + \gamma_i + \epsilon_{ij}, \quad z = 0,1.$$

In the above model, we assume $\gamma_i \sim N(0, \sigma_\gamma^2)$ as a random intercept, and $\epsilon_{ij} \sim N(0, \sigma_\epsilon^2)$ is an independent error term. Note that under this simulation setting, the outcomes are independently and identically distributed conditional on each cluster but are correlated marginally across clusters. The intraclass correlation coefficient is given by $\rho = \sigma_\gamma^2 / (\sigma_\gamma^2 + \sigma_\epsilon^2)$, and chosen to be either 0.01 or 0.1 in our simulation, reflecting a small and moderate intraclass correlation coefficient.[20,21] To specify the other outcome model parameters, we consider two scenarios:

(i) Non-differential outcome models (i.e. homogenous treatment effects) between the always-recruited and compliant-recruited, i.e., $\mu_a = \mu_c$, $\tau_a = \tau_c$ and $\lambda_a = \lambda_c$;

(ii) Differential outcome models between the always-recruited and compliant-recruited, i.e., at least one of the following holds: $\mu_a \neq \mu_c$, $\tau_a \neq \tau_c$, $\lambda_a \neq \lambda_c$.

Numerical specification of these parameters is provided in Table 1 along with the simulation results. We simulate 2,000 trial replicates under each parameter combination. The true value of $\tau^R$ for each scenario is calculated using Result 1 with modifications suggested in Appendix B. In each simulated replicate, we use multivariate adjustment to estimate $\tau^R$, where we fit a linear mixed model by regressing the observed outcome on the cluster treatment indicator and patient-level covariates $X_{ij}$ and take the coefficient of the treatment indicator as the covariate-adjusted ATE estimator for $\tau^R$. We assess the percent bias, precision and coverage rate of the estimator. In each scenario, the true value of $\tau^R$ can differ and is computed via Monte Carlo simulations (also presented in Table 1).

[Table 1 about here]

Two key observations are in order from Table 1. First, under scenario (i), when the outcome models (and thus treatment effects) between the always-recruited and compliant-recruited strata are homogenous, the post-randomization selection can be fully controlled by covariate adjustment. This is demonstrated by the small relative bias, agreement between the Monte Carlo standard deviation and the mean estimated standard error, as well as the nominal coverage in the first four rows of Table 1. In fact, scenario (i) is consistent with the simulation design in Leyrat et al.[13,14] In their simulations, they implicitly assumed a common outcome model across the principal strata, and therefore as expected, their results recommended multivariate (or propensity score) adjustment to remove selection bias in cluster randomized trials. This is a case where covariate adjustment can help remove selection bias in the recruited sample.

Second, under scenario (ii), when the true outcome models differ between the two latent strata for at least one component (e.g., there is an interaction between principal strata membership and the intercept, treatment or covariates), multivariate adjustment leads to significant biases and under-coverage for estimating $\tau^R$, regardless of the magnitude of intraclass correlation coefficient. The more heterogeneous the two strata with respect to the true outcome model, the larger the percent bias of the treatment effect estimator. In scenario (ii), with only the recruited sample, we generally do not have enough information to differentiate between the always-recruited and compliant-recruited in the intervention clusters, and therefore none of the covariate adjustment method is unbiased for estimating $\tau^R$.

We have also conducted simulations that uses propensity score weighting methods[13,14] to adjust for covariate imbalance and the conclusions remain the same as the above.

**Conclusions**

In cluster randomized trials, individual subjects are usually recruited after the initial randomization and the randomization label is open to both recruiters and potential participants.[5,12] As such, post-randomization selection often occurs because the recruitment process can differ between arms. Consequently, the recruited population may not be representative of the overall population, and it is important to differentiate between the ATE estimands defined on the two populations. Moreover, the recruited subjects are often systematically different between the arms. In this paper, we elucidate such post-randomization selection bias via the principal stratification framework, which classifies subjects into subpopulations (i.e., principal strata) based on their joint potential recruitment status under both arms. We analytically express both ATE estimands as weighted averages of different principal causal effects. We also show that post-randomization selection renders the recruited subjects in the treatment and control arm to be composed of different principal strata. When the recruitment process is different between the arms and treatment effects are heterogeneous across principal strata, a naïve ITT analysis of the recruited subjects would usually be biased for both ATEs.

In practice, a common flag of selection bias is imbalance of the baseline covariates between intervention and control arms. We clarify that covariate imbalance caused by post-randomization selection is distinct from chance imbalance. Specifically, because the principal stratum membership is latent and it is usually associated with both treatment status and with the outcome, post-randomization selection essentially induces a type of unmeasured confounding that randomization cannot prevent. Therefore standard covariate adjustment methods—which are designed to adjust for chance imbalance—are generally not sufficient to correct for the imbalance caused by post-randomization selection except for some specific settings.

The recruitment process may be viewed as a missing data generating process, with the data on the non-recruited subjects being "missing." The missingness pattern in our setting is structural, in the sense that the entire never-recruited stratum and the compliant-recruited stratum under the control condition are missing. Therefore, the missing data mechanism is *missing not at random*.[22] This implies that usually we cannot use standard missing data methods such as multiple imputation to address the problem, which only applies to the *missing at random* setting. Unless the treatment effects are homogenous across principal strata, the recruited sample, which consists of always-recruited and compliant-recruited subjects, does not contain information to impute the missing data on the never-recruited subjects. Instead, additional data and estimation strategies are necessary to estimate either the ATE on the overall population $\tau^O$ or on the recruited population $\tau^R$. Specifically, it is necessary to collect outcome data on at least some of the un-recruited patients as well as covariates that are predictive of subjects' participation in a trial. Such data allow us to leverage mixture models to predict each individual's principal stratum membership and then estimate the stratum-specific heterogeneous treatment effects $\tau_s$ and consequently $\tau^O$ and $\tau^R$. There is an extensive literature in causal inference on estimating principal causal effects based on mixture models,[23–27] but these methods have not been applied to the context of recruitment bias in cluster randomized trial, and would require some adaption. A detailed exposition of principal stratification analysis in the setting of recruitment bias in cluster trials is beyond the scope of this paper, which focuses on identifying the problem, and is subject to our ongoing research.

Throughout our discussion, we assume within each cluster the potential outcomes of the individuals are independently and identically distributed. However, interference between individuals within the same cluster often exists, that is, one individual's potential outcome is affected by the treatment assignments to other individuals. This would complicate the definition of estimands and estimation strategies. Despite the emerging literature on interference in the causal inference literature,[28,29] this topic has been rarely discussed in the context of clinical trials and deserves more attention from both trialists and methodologists.

A key question in design is how to obtain the aforementioned additional data from participating clusters.[30] In pragmatic trials, such data may be available from external sources such as electronic health records. Another useful source of information would be baseline covariates that are predictive of the subjects' participation decision. For example, when the inclusion into a study is conducted via recruitment, trialists can add questions during the recruitment like "would you participate in this study had you been assigned to the other arm?" or "what factors affect your decision of participating this study?" Such information helps to predict a subject's principal stratum and thus estimate the principal causal effects.

As a general guideline, in the design stage of a clinical trial, the investigators should routinely assess the possibility of post-randomization selection bias. If the possibility is deemed high, then they should first adopt common recommendations in the literature to reduce such bias through design.[5,31] If selection bias cannot be avoided, as is the case in many cluster trials, then they should consider to collect more data on at least a portion of the non-recruited subjects and covariates that are predictive of patients' recruitment status.[30] Then in the analysis stage, one can conduct a formal principal stratification analysis[15,24] to validly estimate the causal effects. Overall, selection bias is best avoided through careful design than trying to account for it through analysis. It would be prudent for trialists to, before the study, weigh the potential benefits and costs of a cluster randomization design to determine whether it is justified, and if not, consider choosing an individually randomized design.

**Appendix**

**A. Proof of Result 1**.

For simplicity, we drop the subscript in the following derivation. (Removing the subscript also makes it clear that the result can be applied to both cluster and individual randomized trials.)

Under the monotonicity assumption, we have

$$\tau^R = \sum_{s=a,c} \tau_s P(S = s | R = 1). \tag{A.1}$$

Because

$$P(S = s | R = 1) = \sum_{z=0,1} P(S = s | R = 1, Z = z) P(Z = z | R = 1), \tag{A.2}$$

we need to identify the two components $P(S = s | R = 1, Z = z)$ and $P(Z = z | R = 1)$.

*Step 1: identify $P(S = s | R = 1, Z = z)$.*

First, as we elaborated in the main text, it is straightforward to verify that under monotonicity, the recruited subjects in the control arm are all always-recruited, i.e., $P(S = a | Z = 0, R = 1) = 1$, and $P(S = c | Z = 0, R = 1) = 0$.

The intervention arm consists of always-recruited and compliant-recruited. By definition of conditional probability, for each stratum $s$, we have

$$P(S = s|Z = 1, R = 1) = \frac{P(R = 1|S = s, Z = 1) \times P(S = s|Z = 1)}{P(R = 1|Z = 1)} \qquad (A.3)$$

For always-recruited and compliant-recruited, we have $P(R = 1|S = s, Z = 1) = 1$. Also, due to randomization in the overall population, we have $P(S = s|Z = 1) = P(S = s)$. Plugging these two equations into Equation (A.3) and take the ratio between always-recruited and compliers, we obtain

$$\frac{P(S = a|Z = 1, R = 1)}{P(S = c|Z = 1, R = 1)} = \frac{P(S = a|Z = 1)}{P(S = c|Z = 1)} = \frac{P(S = a)}{P(S = c)} \qquad (A.4)$$

(A.4) implies that the ratio between the proportion of always-recruited and compliant-recruited in the overall population is the same as the ratio of the proportion of always-recruited and compliant-recruited in the recruited intervention arm. This helps to identify the marginal probabilities of always-recruited and compliant-recruited in the recruited intervention arm (which adds up to 1) given the marginal probabilities of each strata in the overall population. Specifically, let $p_s = P(S = s)$, then we have $P(S = c|Z = 1, R = 1) = p_c/(p_c + p_a)$, and $P(S = a|Z = 1, R = 1) = p_a/(p_c + p_a)$.

$$P(S = c|Z = 1, R = 1) = \frac{p_c}{p_c + p_a}, \qquad P(S = a|Z = 1, R = 1) = \frac{p_a}{p_c + p_a} \qquad (A.5)$$

*Step 2: identify $P(Z = z|R = 1)$.*

Note that

$$P(Z = z|R = 1) = \frac{P(R = 1|Z = z)P(Z = z)}{P(R = 1)} \qquad (A.6)$$

Because $P(R = 1|Z = z) = \sum_{s=a,c} P(R = 1|S = s, Z = z)P(S = s|Z = z)$, we have

$$P(R = 1|Z = 0) = P(R = 1|S = a, Z = 0)P(S = a|Z = 0) = p_a,$$

$$P(R = 1|Z = 1) = p_a + p_c.$$

Denote the randomization probability as $P(Z = 1) = r$. So, the total recruitment rate is

$P(R = 1) = \sum_{z=0,1} P(R = 1|Z = z)P(Z = z) = (p_a + p_c)r + p_a(1 - r) = p_a + rp_c$. Plugging these expressions into formula (A.6), we have

$$P(Z = 1|R = 1) = \frac{(p_a+p_c)r}{p_a+rp_c}, \quad P(Z = 0|R = 1) = \frac{p_a(1-r)}{p_a+rp_c} \quad (A.7)$$

*Step 3: identify $\tau^R$.*

Plugging (A.5) and (A.7) into (A.2), we obtain the marginal probability of always-recruited and compliant-recruited in the recruited sample to be

$$P(S = c|R = 1) = \frac{rp_c}{rp_c + p_a}, \quad P(S = a|R = 1) = 1 - \frac{rp_c}{rp_c + p_a} \quad (A.8)$$

Plugging (A.8) into (A.1), we prove

$$\tau^R = \frac{rp_c}{rp_c+p_a}\tau_c + \left(1 - \frac{rp_c}{rp_c+p_a}\right)\tau_a. \quad \blacksquare$$

**B. Randomization probability to ensure balanced samples between arms.**

If we want to ensure the sample size of the recruited subjects are similar between two arms, just setting $P(Z = 1|R = 1) = P(Z = 0|R = 1)$. Plugging the expressions in (A.7) into the equation we obtain: $(p_a + p_c)r = p_a(1 - r)$. Solving $r$ gives $r = p_a/(2p_a + p_c)$.

|  Full Data | | | | | Obs Data from a Randomized Study | |
|---|---|---|---|---|---|---|
| Principal Stratum $S$ | Post-random recruitment $R(1)$ | $R(0)$ | Potential outcome $Y(1)$ | $Y(0)$ | Average of $(R,Y)$ given assignment $Z=1$ | $Z=0$ |
| Always | 1 | 1 | 30 | 10 | (1, 27.5) | (1, 10) |
| Compliant | 1 | 0 | 25 | 10 |  | (0, ?) |
| Never | 0 | 0 | 20 | 10 | (0, ?) |  |

**Figure 1**: *In this hypothetical example of a cluster clinical trial, we assume (i) monotonicity, (ii) there are no defiant-recruited, and (iii) equal proportions of each stratum in the overall population. We drop subscripts for simplicity. Here "?" means that the corresponding data are not observed because the subjects are not recruited. The left part shows the full data, and the right part shows the observed data, which is a subset of the full data.*

***Table 1.*** *True value of ATE ($\tau^R$), percent bias, Monte Carlo standard deviation (MCSD), mean estimated standard error (ESE), and coverage probability (CP) of the 95% confidence interval of the treatment effect estimator, under different specifications of true data generating models. Scenarios include: (i) non-differential outcome models between the always-recruited and compliant-recruited, i.e., $\mu_a = \mu_c$, $\tau_a = \tau_c$, and $\lambda_a = \lambda_c$; (ii) Differential outcome models between the always-recruited and compliant-recruited, i.e., at least one of the following holds: $\mu_a \neq \mu_c$, $\tau_a \neq \tau_c$, $\lambda_a \neq \lambda_c$.*

|   | True outcome model coefficients | | | | | | Estimand | ICC | Performance metrics | | | |
|---|---|---|---|---|---|---|---|---|---|---|---|---|
|   | $\mu_a$ | $\mu_c$ | $\tau_a$ | $\tau_c$ | $\lambda_a$ | $\lambda_c$ | $\tau^R$ | $\rho$ | % Bias | MCS | ESE | CP |
| (i) | 1.0 | 1.0 | 0.2 | 0.2 | (0.1, 0.1) | (0.1, 0.1) | 0.20 | 0.01 | 0.02 | 0.078 | 0.075 | 95.4 |
|  |  |  |  |  |  |  |  | 0.1 | 1.36 | 0.152 | 0.150 | 94.3 |
|  | 2.0 | 2.0 | 0.8 | 0.8 | (0.2, 0.3) | (0.2, 0.3) | 0.80 | 0.01 | 0.01 | 0.078 | 0.075 | 95.4 |
|  |  |  |  |  |  |  |  | 0.1 | 0.34 | 0.152 | 0.150 | 94.3 |
| (ii) | 1.0 | 2.0 | 0.2 | 0.2 | (0.1, 0.1) | (0.1, 0.1) | 0.20 | 0.01 | 215.03 | 0.081 | 0.078 | 0.1 |
|  |  |  |  |  |  |  |  | 0.1 | 216.41 | 0.153 | 0.151 | 20.2 |
|  | 2.0 | 1.0 | 0.8 | 0.8 | (0.2, 0.3) | (0.2, 0.3) | 0.80 | 0.01 | 53.77 | 0.081 | 0.080 | 0.1 |
|  |  |  |  |  |  |  |  | 0.1 | 53.42 | 0.153 | 0.152 | 21.4 |
|  | 1.0 | 1.0 | 0.2 | 0.8 | (0.1, 0.1) | (0.1, 0.1) | 0.33 | 0.01 | 39.15 | 0.079 | 0.076 | 62.5 |
|  |  |  |  |  |  |  |  | 0.1 | 39.99 | 0.152 | 0.150 | 84.6 |
|  | 2.0 | 2.0 | 0.8 | 0.2 | (0.2, 0.3) | (0.2, 0.3) | 0.67 | 0.01 | 19.22 | 0.079 | 0.077 | 63.2 |
|  |  |  |  |  |  |  |  | 0.1 | 18.81 | 0.152 | 0.151 | 86.4 |
|  | 1.0 | 2.0 | 0.2 | 0.8 | (0.1, 0.1) | (0.1, 0.1) | 0.33 | 0.01 | 169.82 | 0.086 | 0.083 | 0 |
|  |  |  |  |  |  |  |  | 0.1 | 170.66 | 0.156 | 0.153 | 5.5 |
|  | 1.0 | 2.0 | 0.8 | 0.2 | (0.2, 0.3) | (0.2, 0.3) | 0.67 | 0.01 | 44.89 | 0.078 | 0.076 | 3.9 |
|  |  |  |  |  |  |  |  | 0.1 | 45.30 | 0.152 | 0.150 | 48.9 |
|  | 1.0 | 2.0 | 0.2 | 0.8 | (0.1, 0.1) | (0.2, 0.3) | 0.33 | 0.01 | 180.59 | 0.088 | 0.084 | 0 |
|  |  |  |  |  |  |  |  | 0.1 | 181.43 | 0.157 | 0.154 | 4.0 |
|  | 1.0 | 2.0 | 0.8 | 0.2 | (0.2, 0.3) | (0.1, 0.1) | 0.67 | 0.01 | 39.61 | 0.079 | 0.076 | 8.8 |
|  |  |  |  |  |  |  |  | 0.1 | 40.02 | 0.152 | 0.150 | 57.4 |